\documentclass[twocolumn,prl]{revtex4}
\usepackage{amsmath,amssymb,bm}
\usepackage{graphicx}
\usepackage{epstopdf}
\usepackage{latexsym}
\usepackage{subfigure}
\usepackage[usenames,dvipsnames]{color}
\usepackage{hyperref}
\usepackage{natbib}
\usepackage{moreverb}

\begin{document}
\newcommand{\MF}{CH$_3$F}

\newcommand{\todo}[1]{{\color{red}#1}}

\title{Realizing topological states with polyatomic symmetric top molecules }

\author{M. L. Wall$^{1,2}$\footnote{e-mail: mwall.physics@gmail.com},  K. Maeda$^{2}$, and L. D. Carr$^{2}$}

\affiliation{$^{1}$JILA, NIST, Department of Physics, University of Colorado, Boulder, CO 80309, USA}
\affiliation{$^{2}$Department of Physics, Colorado School of Mines, Golden, Colorado 80401, USA}

\begin{abstract}

We demonstrate that ultracold polyatomic symmetric top molecules, such as methyl fluoride, loaded into an optical lattice and subject to {DC electric and microwave} field dressing, can display topological order via a self-consistent analog of {a} proximity effect in the internal state space of the molecule.  The non-trivial topology arises from pairwise transitions between internal states induced by dipole-dipole interactions and made resonant by the field dressing.  Topological order is explicitly demonstrated by matrix product state simulations on 1D chains.  Additionally, we show that in the limit of pinned molecules our description maps onto a long-range and anisotropic XYZ spin model, where Majorana fermions are zero-energy edge excitations in the case of nearest-neighbor couplings. 
\end{abstract}

\maketitle

There is currently a strong push to create systems which harbor Majorana fermion excitations~\cite{Alicea,Mebrahtu}.  The interest in {Majorana fermions} stems from their robustness against local perturbations and connections between {Majorana fermions} and the topology of quantum phases~\cite{NayakRMP,Sachdev,CarrBook}.  The simplest system~\cite{Kitaev} where {Majorana fermions} arise as low-energy edge excitations is in a 1D lattice model of spinless fermions with both tunneling and $p$-wave pairing.  Pairing of fermions is a crucial ingredient for realizing {Majorana fermions}, as it breaks the U(1) symmetry corresponding to conservation of total particle number down to the $\mathbb{Z}_2$ symmetry associated with fermionic parity~\cite{Sau_Halperin,Fidowski_Lutchyn}.  In a solid state setting, $p$-wave pairing in a semiconductor wire can be induced by allowing Cooper pairs to tunnel to the wire from a nearby $p$-wave superconducting reservoir; this is the proximity effect.  In this Letter, we show that a self-induced analog of the proximity effect and other rich features of quantum wire models can be observed with a field-dressed ultracold gas of symmetric top molecules  in which internal states of the molecule provide the discrete wire degrees of freedom.

\begin{figure}[bp]
\centerline{\includegraphics[width=\columnwidth]{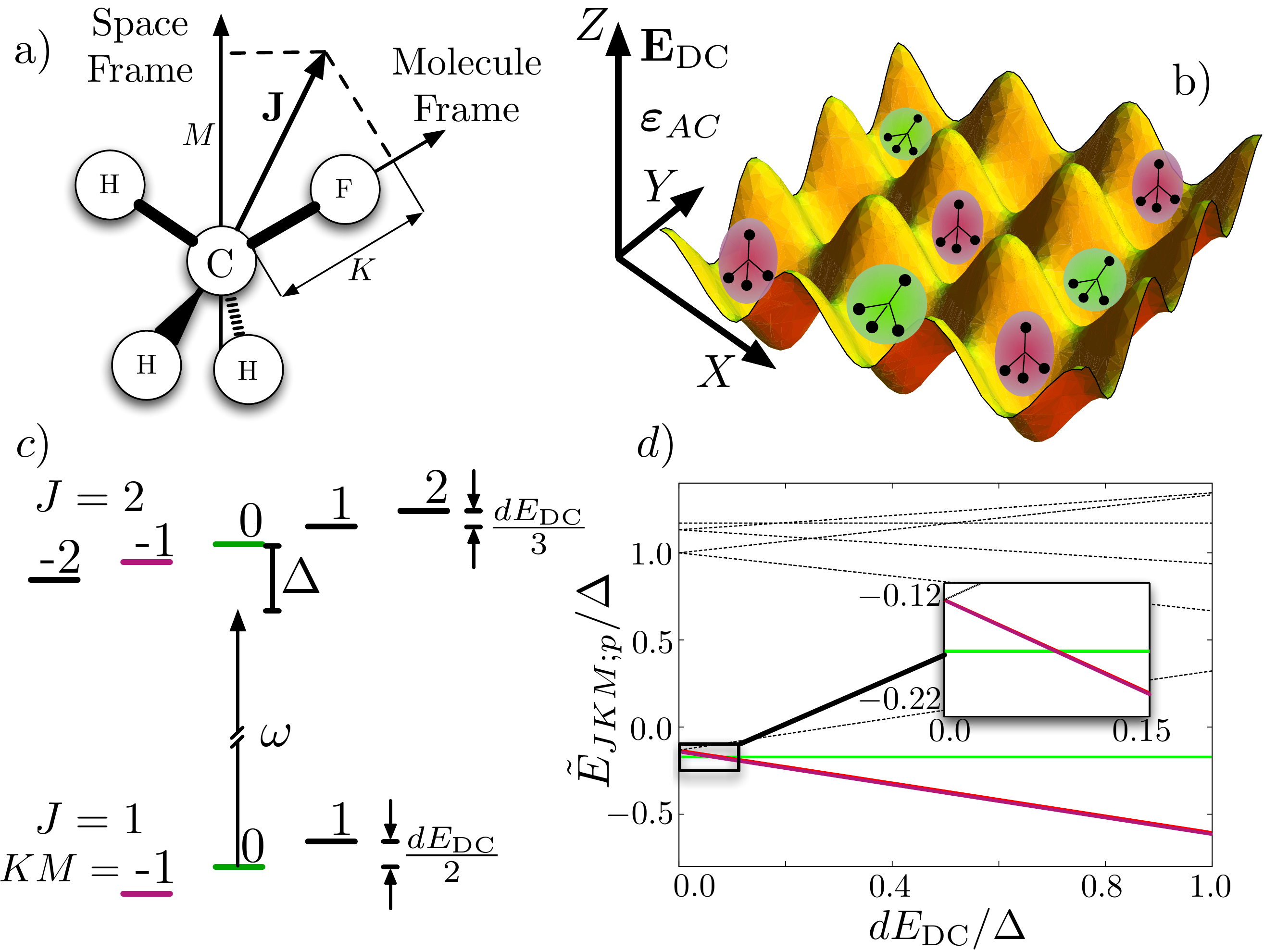}}
\caption{\label{fig:Levels} 
\emph{Dressed symmetric top molecules (STMs) in optical lattices} (a) Rotational angular momentum geometry of the STM {\MF}.  (b) Schematic of field and lattice geometry.  Purple and green denote two internal states.  (c) Levels $|J,K,M\rangle$ and $|J+1,K,M\rangle$ have different Stark effects.  Coupling by a microwave of frequency $\omega$ and detuning $\Delta$ generates dressed state $|\bar{0}\rangle$ ($|\bar{1}\rangle$) in green (purple).  (d) Combined DC and AC Stark shift cause a level crossing of $|\bar{0}\rangle$ and $|\bar{1}\rangle$ ($\Omega/\Delta=1$).  Inset: enlargement of crossing.}
\end{figure}

Polyatomic symmetric top molecules (STMs), of which methyl fluoride, {\MF}, is a canonical example, display a {linear response} to an externally applied static electric field in contrast to the quadratic {response} exhibited by $\Sigma$-state molecules such as the alkali metal dimers~\cite{Wall_Maeda_13}.  The difference arises from a nonzero projection of the rotational angular momentum on the body frame of the molecule, see Fig.~\ref{fig:Levels}(a).  The strong coupling of STMs to external fields is the basis of opto-electrical cooling, a novel route to bring generic STMs to quantum degeneracy~\cite{CH3FOEC}, and also enables STMs to simulate the physics of magnetic dipoles and quantum magnetism with greatly enhanced dipolar interaction energies~\cite{Wall_Maeda_13}.  In addition, STMs have a long history in the physics of molecular beams~\cite{F_H}.  The main idea of our proposal is to use the linear Stark effect for STMs together with a monochromatic microwave field tuned near-resonant to neighboring rotational levels to engineer energy level crossings.  Near such a level crossing, the dipole-dipole interaction causes resonant transitions between the internal states of two molecules.  These transitions preserve the total number of molecules while changing the number of molecules in each internal state by two.  In appropriate geometries, interactions which change the internal state of only one molecule are suppressed, and thus we find an effective many-body model of two-component particles which conserves the total number of particles but explicitly breaks the U(1) symmetry associated with conservation of the relative number of particles in the two internal states down to $\mathbb{Z}_2$~\cite{Cheng_Tu_11}.  Using matrix product state simulations~\cite{Schollwoeck_11}, we explicitly demonstrate topological order in the ground state.  Further, we show that when the translational motion of the molecules is frozen, the dynamics of the rotational excitations can be mapped to a long-range XYZ spin model, where {Majorana fermion} physics is known to exist for the short-range case~\cite{Kitaev,Sela}.

There have been many proposals to realize {Majorana fermions} and other topological phenomena using ultracold gases, including light-induced effective spin-orbit interactions~\cite{Sato,Liu_Jiang,Wang_etal,Cheuk,Spielman}, quantum simulation of the proximity effect using Raman coupling of fermions to a 3D molecular Bose-Einstein condensate~\cite{Jiangetal}, and light- and interaction-induced spatial pair tunneling~\cite{KZ}.  In addition, diatomic molecules have been proposed to generate topological phases~\cite{Micheli_topo,EGDT,Cooper_Shlyapnikov,Gorshkov}.  Two important distinctions of our work from others are that pair tunneling occurs between internal states of the molecule rather than in real space~\cite{SyntheticDims}, and pair tunneling and interactions are anisotropic and long-range, arising from the dipole-dipole interaction.  In addition to the fundamental interest of a quantum mechanical object providing pairing fluctuations for itself, the former distinction is also key for detection; in ultracold gases it is vastly easier to perform measurements on specific internal states than on localized points in space.

\emph{Field dressing of symmetric top molecules.}  The rotational degrees of freedom of a STM in the lowest electronic and vibrational state may be characterized by the basis $|J,K,M\rangle=\sqrt{\frac{2J+1}{8\pi^2}}\mathcal{D}^{J\ast}_{MK}(\omega_m)$, where $J$ is the rotational quantum number, $M$ is the projection of rotation $\mathbf{J}$ on a space-fixed quantization axis, $K$ is the projection of $\mathbf{J}$ on the symmetry axis of the molecule, and $\mathcal{D}^{J}_{MK}(\omega_m)$ are the matrix elements of the Wigner $D$-matrix rotating the space-fixed frame to the molecule-fixed frame by the Euler angles $\omega_m$~{\cite{Zare_1988}}, see Fig.~\ref{fig:Levels}(a).  The corresponding rotational eigenenergies are $E_{JKM}=B_0J(J+1)+(A_0-B_0)K^2$, where the rotational constants $B_0\approx25$GHz, $A_0\approx 155$GHz for {\MF}.  In a static electric field of strength $E_{\mathrm{DC}}\ll B_0/d$ defining the quantization axis, with $d$ the permanent dipole moment, the matrix elements of the dipole operator along space-fixed spherical direction $p$, $\hat{d}_p$, take the form of a spherical tensor with reduced matrix element $\langle J,K'||\hat{\mathbf{d}}||J,K\rangle=dK\sqrt{\frac{2J+1}{J(J+1)}}\delta_{K,K'}$~\cite{Wall_Maeda_13}.  Hence, STMs in this field regime display a linear Stark effect with eigenenergies {$E_{JKM}=dKME_{\mathrm{DC}}/[J(J+1)]$}, as shown in Fig.~\ref{fig:Levels}(c).  The effects of hyperfine structure are considered later in this work.

To understand the microwave dressing procedure, let us now consider applying a microwave field $\mathbf{E}_{\mathrm{AC}}$ with linear polarization along the space-fixed quantization axis, $\boldsymbol{\varepsilon}_{\mathrm{AC}}=\mathbf{e}_Z$, see Fig.~\ref{fig:Levels}(b), which is red-detuned an amount $\Delta$~\cite{hbarfootnote} from resonance with the $|J,K,0\rangle\to|J+1,K,0\rangle$ transition, as shown in Fig.~\ref{fig:Levels}(c).  Applying the rotating wave approximation and transforming to the Floquet picture~\cite{Shirley}, the quasienergies are obtained by solving the Schr\"odinger equation for fixed $M$ with the $2\times 2$ Hamiltonians~\cite{tbtfootnote}:
\begin{align}
\label{eq:HJKM}\hat{H}_{JKM}&=\left(\begin{array}{cc} -\frac{dKME_{\mathrm{DC}}}{J\left(J+1\right)}&-\Omega_{JKM}\\ -\Omega_{JKM}&\Delta-\frac{dKME_{\mathrm{DC}}}{\left(J+1\right)\left(J+2\right)}\end{array}\right)\, ,
\end{align}
where $\Omega_{JKM}\equiv \Omega\{\frac{[(J+1)^2-K^2][(J+1)^2-M^2]}{(J+1)^2(2J+1)(2J+3)}\}^{1/2}$ with the Rabi frequency $\Omega\equiv dE_{\mathrm{AC}}$.  Single-particle eigenstates of Eq.~\eqref{eq:HJKM} in the rotating frame will be denoted by an overbar, e.g.~, $|\bar{0}\rangle$.

In the perturbative regime where $\Omega,dE_{\mathrm{DC}}\ll \Delta$, the quasienergies are split into manifolds $\tilde{E}_{JKM;\pm}$ separated by roughly $\Delta$, see Fig.~\ref{fig:Levels}(d).  The $M$ dependence of the off-diagonal components $\Omega_{JKM}$ introduces an effective tensor shift between states of different $M$ which is proportional to $\Omega^2$, similar to the microwave-induced quadratic Zeeman effect in spinor Bose gases~\cite{MIQZ}.  Including the static field $E_{\mathrm{DC}}$ can cause two such quasienergy levels with different $M$ to cross as the static field energy $dE_{\mathrm{DC}}$ becomes of the order of the effective tensor shift, as shown for the case of the $(J,K)=(1,1)\to(2,1)$ transition in Fig.~\ref{fig:Levels}(c).  Note that level crossings can also be engineered outside of the perturbative regime.  We will denote the parametric relationship of the Rabi frequency and the electric field at such a crossing as $\tilde{\Omega}\left(E_{\mathrm{DC}}\right)$.

The components of the dressed states in the $|J+1,K,M\rangle$ manifold oscillate in time with frequency $\omega$.  Hence, the dipole moments of the dressed states contain both static and time-oscillating pieces.  While the oscillating terms time-average to zero for a single molecule, the dipole-allowed exchange of rotational quanta for two molecules can be resonant due to the two dipoles oscillating in phase~\cite{Cooper_Shlyapnikov,Yan_Moses}.  Maintaining a single frequency for the microwave field but allowing for different polarizations and intensities realizes vast tunability over the various interaction processes.  The only assumption we use in this work is that the dipole moments of two states near a level crossing only have static components along a single space-fixed spherical direction.  Practically, the microwave field can contain either $p=\pm 1$ components or $p=0$ components, but not both.  The requirement of only a single microwave frequency  is in contrast to proposals with $^1\Sigma$ molecules, which often require precise frequency and polarization control of {multiple microwaves} in order to realize topological phases~\cite{Gorshkov}.

\emph{Effective many-body model near a level crossing.}  We now consider an ensemble of fermionic STMs (e.g.~$^{13}$CH$_3$F or CH$_3$CN) trapped in an optical lattice with a quasi-2D geometry and prepared near a level crossing with $\mathbf{E}_{\mathrm{DC}}$ normal to the plane of molecules, see Fig.~\ref{fig:Levels}(b).  Our arguments also apply for a quasi-1D arrangement consisting of one row of Fig.~\ref{fig:Levels}(b).  Two molecules interact through the dipole-dipole interaction
\begin{align}
\label{eq:HDD}\hat{H}_{\mathrm{DD}}&=\textstyle-\sqrt{6}\sum_{p=-2}^{2}(-1)^pC^{(2)}_{-p}(\mathbf{R})[\hat{\mathbf{d}}\otimes \hat{\mathbf{d}}]^{(2)}_{p}/R^3\, ,
\end{align}
where $C^{(2)}_p=\sqrt{\frac{4\pi}{5}}Y^{(2)}_p(\mathbf{R})$ is an unnormalized spherical harmonic, $\mathbf{R}$ is the relative coordinate with $R$ its magnitude, and $[\hat{\mathbf{d}}\otimes \hat{\mathbf{d}}]^{\left(2\right)}_{p}$ is the $p^{\mathrm{th}}$ component of the rank-two tensor product of dipole operators in the space-fixed spherical basis~\cite{Zare_1988}.  The terms in Eq.~\eqref{eq:HDD} with $p=\pm 1$ are proportional to $\sin\theta\cos\theta$, where $\theta$ is the polar angle between $\mathbf{R}$ and $\mathbf{E}_{\mathrm{DC}}$.  For the geometry in Fig.~\ref{fig:Levels}(b), $\theta=\pi/2$ and so these terms vanish.  The remaining interactions are the $p=0$ term, $(1-3\cos^2\theta)[\hat{d}_0\hat{d}_0+(\hat{d}_1\hat{d}_{-1}+\hat{d}_{-1}\hat{d}_1)/2]/R^3$, and the $p=\pm2$ terms, $-3\sin^2\theta e^{-2i\phi}\hat{d}_1\hat{d}_1/(2R^3)+\mathrm{h.c.}$, where $\phi$ is the angle between $\mathbf{R}$ and the space-fixed $x$ axis in the plane of molecules.

\begin{figure}[tbp]
\centerline{\includegraphics[width=1.0\columnwidth]{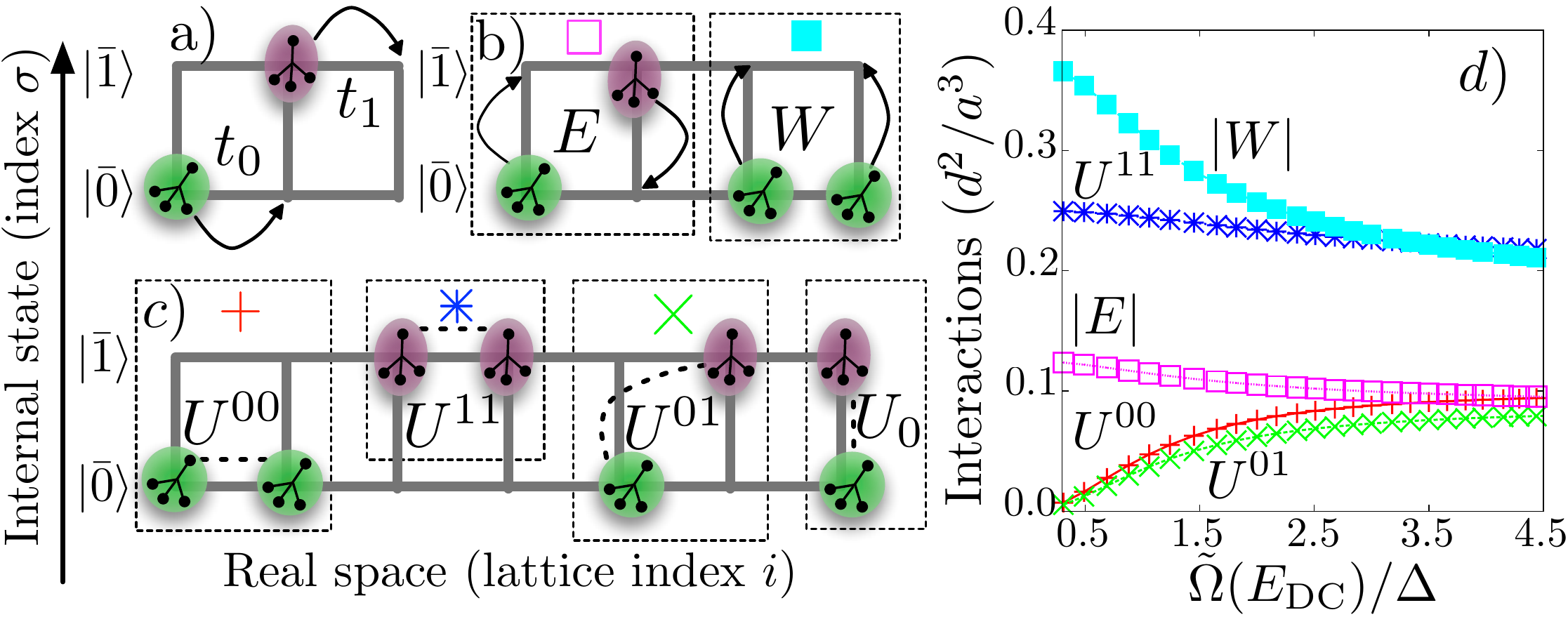}}
\caption{\label{fig:ID}  \emph{Interaction processes in the effective lattice Hamiltonian.} The two internal states $|\bar{0}\rangle$ and $|\bar{1}\rangle$ may be viewed as an discrete spatial degree of freedom, e.g.~a ladder. (a) Tunneling rates $t_{\sigma}$ depend on the internal state due to polarizability anisotropy~\cite{Wall_Maeda_13,Neyenhuis_Yan_12}.  (b) $E$ and $W$ interactions change the internal state of the molecules.  $E$ processes preserve the number in each internal state, $W$ processes change it by $\pm 2$.  (c) $U$ interactions preserve the internal state of the molecules.   d) Nearest-neighbor Hubbard parameters ($W\equiv W_{i,i+1}$ etc.) of the many-body model Eq.~\eqref{eq:MBModel} with ($E_{\mathrm{DC}}$-dependent) Rabi frequency $\tilde{\Omega}$ at the level crossing in Fig.~\ref{fig:Levels}(d).  Symbols correlate processes to Panels (a)-(c).
}
\end{figure}
Expanding the field operator in a basis of lowest band Wannier functions $w_{i\sigma}(\mathbf{r})$~\cite{Jaksch,Lewenstein}, where $\sigma\in \left\{0,1\right\}$ indexes the dressed states $\left\{|\bar{0}\rangle,|\bar{1}\rangle\right\}$ which are nearly resonant, we find the lattice Hamiltonian
\begin{align} \label{eq:MBModel}
\nonumber \hat{H}&=-\textstyle\sum_{\langle i,j\rangle\sigma}t_{\sigma}\hat{a}_{i\sigma}^{\dagger}\hat{a}_{j\sigma}+\delta \sum_{i}\hat{n}_{i1}\\
\nonumber &-\textstyle \frac{1}{2}\sum_{i,j,i\ne j}[E_{i,j}\hat{S}^+_{i}\hat{S}^{-}_j+W_{i,j}\hat{S}^+_i\hat{S}^+_j+\mathrm{h.c.}]\\
&+\frac{1}{2}\sum_{\sigma,\sigma', i,j,i\ne j}U^{\sigma\sigma'}_{i,j}\hat{n}_{i\sigma}\hat{n}_{j\sigma}+\textstyle \sum_{i}U_0\hat{n}_{i0}\hat{n}_{i1}\, .
\end{align}
Here, $\hat{a}_{i\sigma}$ destroys a STM in Wannier state $w_{i\sigma}(\mathbf{r})$, $\hat{n}_{i\sigma}=\hat{a}_{i\sigma}^{\dagger}\hat{a}_{i\sigma}$, and $\hat{S}^+_i=\hat{a}_{i0}^{\dagger}\hat{a}_{i1}$, $\hat{S}^-_i=(\hat{S}_i^+)^{\dagger}$ are the usual spin-1/2 ladder operators.  In order, the terms in Eq.~\eqref{eq:MBModel} are state-dependent tunneling $t_{\sigma}$ of molecules between neighboring lattice sites $\langle i,j\rangle$; a single-particle energy offset $\delta$ of state {$|\bar{1}\rangle$} with respect to state $|\bar{0}\rangle$; state-exchanging collisions $E_{i,j}$ of molecules at sites $i$ and $j$; state-transferring collisions $W_{i,j}$ which transform two molecules in state $|\bar{0}\rangle$ at sites $i$ and $j$ into the state $|\bar{1}\rangle$ and vice versa;  state-preserving collisions $U_{i,j}^{\sigma\sigma'}$ between molecules in states $\sigma$ and $\sigma'$ at lattice sites $i$ and $j$, respectively;  and on-site interactions $U_{0}$.  A schematic view of the processes in Eq.~\eqref{eq:MBModel} is given in Fig.~\ref{fig:ID}(a)-(c).  The magnitudes of the Hubbard parameters for the specific level crossing in Fig.~\ref{fig:Levels}(d) are displayed in Fig.~\ref{fig:ID}(d) as a function of the $E_{\mathrm{DC}}$-dependent Rabi frequency at the level crossing, $\tilde{\Omega}(E_{\mathrm{DC}})$. For the dressing scheme in Fig.~\ref{fig:Levels}(c), the Hubbard parameters $U$ and $E$ are overlaps of the $p=0$ component of the dipole-dipole potential in the basis of Wannier functions~\cite{Wall_Carr_CE}, while the $W$ terms involve overlaps of the $p=\pm 2$ components of the dipole-dipole potential.  All dipolar parameters $U$, $E$, and $W$ have an approximately $1/|i-j|^3$ decay between lattice sites, and the $W$ terms additionally feature a dependence on the angle $\phi$.  Other dressing schemes divide the angular dependence between $U$, $E$, and $W$.

Because our {scheme populates multiple dressed states consisting of different rotational levels}, molecules {undergo possibly rapid rotationally inelastic processes at short range} which will cause a loss of molecules from the trap.  In many cases the on-site interaction $U_0$ is large and positive, enforcing a hard-core constraint.  However, even in the cases of $U_0\le 0$, molecules may be forbidden from occupying the same lattice site by the quantum Zeno effect.  The quantum Zeno effect has been shown to enforce a hard-core constraint for KRb, where {two-body losses} are due to chemical reactions, and gives rise to lifetimes which are long compared to the typical time scales of interactions~\cite{Yan_Moses,Zeno}.  The numerical results given in this work have $U_0>0$ large and so should not be affected by rotationally inelastic losses.

\emph{Many-body features of the effective Hamiltonian} The Hamiltonian Eq.~\eqref{eq:MBModel} has a U(1) symmetry generated by the total number operator $\hat{N}=\hat{N}_0+\hat{N}_1$ with $\hat{N}_{\sigma}= \sum_{i}\hat{n}_{i\sigma}$.  The $W$ term breaks number conservation within each internal state, but preserves the parities defined by $\hat{P}_{\sigma}=\exp(-i\pi \hat{N}_{\sigma})$.  Due to the U(1) symmetry, the two parities are redundant, both being proportional to $\hat{P}=\exp[-\frac{i\pi}{2}(\hat{N}_0-\hat{N}_1)]$, which is the parity of the number difference between internal states.  Hence, the internal symmetry of the model Eq.~\eqref{eq:MBModel} is U(1)$\times \mathbb{Z}_2$~\cite{Cheng_Tu_11}.  We can interpret the $W$ term as being a hopping of pairs between two quantum wires or layers, where the wire indices correspond to the dressed states of the molecule, see Fig.~\ref{fig:ID}.  This can be viewed as self-induced analog of the proximity effect, in which the two degenerate dressed states of the molecule resonantly exchange pairs with each other.  Due to the fact that exchange of rotational quanta only occurs when the dipoles oscillate in phase and the particular geometry, dipolar excitation of a single molecule is forbidden.  Single excitation processes which break the $\mathbb{Z}_2$ symmetry can be included systematically by other choices of geometry or field polarization.

In ultracold gases it is often easier to achieve low temperatures for the internal degrees of freedom even when the motional degrees of freedom remain hot.  Hence, a natural first step for many-body physics is to {freeze} the motional degrees of freedom by loading into a deep optical lattice and consider the dynamics of only the internal degrees of freedom~\cite{Yan_Moses}.  In the limit in which the quasi-2D confinement is so deep that the tunneling is negligible, Eq.~\eqref{eq:MBModel} becomes a long-range and anisotropic spin model
\begin{align}
\nonumber \hat{H}&=\textstyle\frac{1}{2}\sum_{i,j,i\ne j}\Big[\left(E_{i,j}+W^{\mathcal{R}}_{i,j}\right)\hat{S}^x_i\hat{S}^x_j+\left(E_{i,j}-W^{\mathcal{R}}_{i,j}\right)\hat{S}^y_i\hat{S}^y_j\\
\nonumber&\textstyle-W^{\mathcal{I}}_{i,j}(\hat{S}^x_i\hat{S}^y_j+\hat{S}^y_i\hat{S}^x_j)+\left(U^{00}_{i,j}+U^{11}_{i,j}-2U^{01}_{i,j}\right)\hat{S}^z_i\hat{S}^z_j\Big]\\
\label{eq:XYZ}&\textstyle+\sum_ih_i\hat{S}^z_i\, ,
\end{align}
where $h_i=\delta+\frac{1}{4}\sum_{j,j\ne i}\left(U^{00}_{i,j}-U^{11}_{i,j}\right)\sum_{\sigma}\hat{n}_{j\sigma}$ is the effective magnetic field at site $i$, we have ignored a constant term, and $W^{\mathcal{R}}_{i,j}$ ($W^{\mathcal{I}}_{i,j}$) is the real (imaginary) part of $W_{i,j}$.  Note that Eq.~\eqref{eq:XYZ} does not conserve magnetization {due to the non-zero $W^{\mathcal{R}}$ and $W^{\mathcal{I}}$ terms easily accessible in our scheme}, in contrast to the XXZ models realized with alkali dimer molecules~\cite{Gorshkov_Manmana,Yan_Moses}.  In one dimension, choosing coordinates such that $W^{\mathcal{I}}_{i,j}=0$, and considering that the coefficient of $\hat{S}^z_i\hat{S}^z_j$ vanishes~\cite{vanishfootnote}, Eq.~\eqref{eq:XYZ} becomes a long-ranged version of the XY model in a longitudinal field.  The nearest-neighbor XY model is equivalent to the Kitaev wire Hamiltonian~\cite{Kitaev}, where Majorana fermions are known to exist.  It was also pointed out that long-range interactions 
may not qualitatively change the nature of topological phases~\cite{vanishfootnote}.  Finally, we note that in the limit of motionally quenched molecules, the statistics are unimportant, and so one can also realize Eq.~\eqref{eq:XYZ} with a bosonic STM, such as $^{12}$CH$_3$F or the other methyl halides.

To demonstrate non-trivial topology in the ground state of Eq.~\eqref{eq:MBModel} in one dimension, we compute the entanglement splitting $\Delta\lambda=\sum_{i}\left(\lambda_{2i+1}-\lambda_{2i}\right)$, where $\lambda_i$ is the $i^{\mathrm{th}}$ Schmidt value of a bipartite splitting in the center of the 1D chain, and the gap between the even and odd fermionic parity sectors using variational matrix product state algorithms~\cite{Schollwoeck_11}.  For the dressing scheme of Fig.~\ref{fig:Levels}(d) with $\tilde{\Omega}(E_{\mathrm{DC}})/\Delta\approx 1.5$, lattice filling $N/L=2/3$, and tunneling in the $|\bar{0}\rangle$ state $t_0=0.1W$, we find a vanishing of the entanglement splitting near $\delta\approx-0.6 W$, where the gap between the even and odd parity sectors also closes.  The vanishing of $\Delta\lambda$ explicitly demonstrates the topological order~\cite{ES}.  We find qualitatively similar behavior of $\Delta \lambda$ for a variety of more complex dressing schemes and tunnelings. 

 In our analysis of the field dressing of STMs we have neglected hyperfine structure.  Though the hyperfine structure of STMs is complicated~\cite{Wall_Maeda_13}, a single hyperfine component may be selected via a strong magnetic field, similarly to the alkali dimers~\cite{Hyperfineselection}.  Alternatively, working at microwave detuning large compared to the typical hyperfine splittings, $\Delta\gg E_{\mathrm{hfs}}\approx 10$kHz for {\MF}, one can address all hyperfine states equally with a readily achievable microwave power on the order of tens of W/cm$^2$.  While we have focused on polyatomic STMs in which all states with a given $J$ and $K$ are degenerate in zero DC field, we expect similar level crossings in other systems with a linear Stark effect but no zero-field degeneracy, such as the Lambda doublet of OH~\cite{OH}, its fermionic analog OD~\cite{OD}, or other species with non-zero projection of orbital angular momentum along the body frame $|\Lambda|>0$.

We have found a general mechanism for generating level crossings between internal states with a finite transition dipole matrix element in symmetric top molecules by a combination of microwave dressing and the linear Stark effect.  The dipole-dipole interaction generates resonant pair transitions between such nearly degenerate levels.  By appropriate choices of geometry and field polarization, transfer of a single molecule between internal states can be forbidden, and the resulting many-body system features topologically nontrivial states.  Our results provide a new route towards the study of topological phases in many-body physics by harnessing the rich internal structure of molecules.

We acknowledge useful conversations with Ryan Mishmash and Christina Kraus.  This work was supported by {the AFOSR under grants FA9550-11-1-0224 and FA9550-13-1-0086}, {ARO grant number 61841PH, ARO-DARPA-OLE}, and the National Science Foundation under Grants PHY-1207881, PHY-1067973,  PHY-0903457, {PHY-1211914, PHY-1125844}, and NSF PHY11-25915.  We also acknowledge the Golden Energy Computing Organization at the Colorado School of Mines for the use of resources acquired with financial assistance from the National Science Foundation and the National Renewable Energy Laboratories.  We would also like to thank the KITP for hospitality.

\newpage


\begin{thebibliography}{10}
\bibitem{tbtfootnote}
States in the $(J+1,K)$-manifold which do not couple to any states in the $\left(J,K\right)$-manifold have trivial $1\times 1$ Hamiltonians.  These states do not produce pair hopping.
\bibitem{vanishfootnote}
The coefficient of $\hat{S}^z_i\hat{S}^z_j$ can be made to vanish for a continuous range of Rabi frequencies with a single microwave frequency; this does not require fine-tuning to a specific point in parameter space.
\end{thebibliography}
\end{document}